\documentclass{pnastwo}

\newcommand{\bea}{\begin{eqnarray}}
\newcommand{\eea}{\end{eqnarray}}

\usepackage{amssymb,amsfonts}
\usepackage{cite}
\usepackage{color}
\usepackage[fleqn]{amsmath}
\usepackage[dvips]{graphicx}
\usepackage{subfigure}
\begin{document}

\title{A Unifying Decomposition and Reconstruction Model for Discrete Signals}

\author{Yiguang Liu\affil{1}{Computational Imaging and Multidimensional Processing Lab, College of Computer Science, Sichuan University, Chengdu, Sichuan
Province, China, 610064}}

\maketitle

\begin{article}
\begin{abstract}
Decomposing discrete signals such as images into components is vital in many applications, and this paper propose a framework to produce filtering banks to accomplish this task. The framework is an equation set which is ill-posed, and thus have many solutions. Each solution can form a filtering bank consisting of two decomposition filters, and two reconstruction filters. Especially, many existing discrete wavelet filtering banks are special cases of the framework, and thus the framework actually makes the different wavelet filtering banks unifiedly presented. Moreover, additional constraints can impose on the framework to make it well-posed, meaning that decomposition and reconstruction (D\&R) can consider the practical requirements, not like existing discrete wavelet filtering banks whose coefficients are fixed. All the filtering banks produced by the framework can behave excellently, have many decomposition effect and precise reconstruction accuracy, and this has been theoretically proved and been confirmed by a large number experimental results.
\end{abstract}

\keywords{Discrete wavelet transform| Filtering banks| Decomposition and reconstruction (D\&R)| Framework}


\section{Significance Statement}
This paper has contributed that
\begin{description}
  \item[i:] An equation set model is proposed, and any solution of model can act as the coefficients of D\&R filters;
  \item[ii:] Existing discrete wavelet filters (such as Daubechies, Coiflets, Symlets, Meyer, etc) are solutions of the model, and simultaneously many other D\&R filters can get from the model.
  \item[iii:] Special constraints can be additional imposed on the model, making the D\&R filters applicable to flexible practical requirements contrasted with existing wavelet filters whose coefficients are fixed.
\end{description}
In all, a unifying D\&R model is proposed with discrete wavelet decomposition being special cases , and has wide applications in image processing, time series analysis, etc.
\section{Introduction}
\dropcap{D}ecomposing signals into components are very useful in data compression, feature discrimination or even fractal analysis, touching upon image processing, machine learning, big data analysis, artificial intelligence, etc \cite{1}\cite{2}\cite{3}\cite{4}. Actually, we can get more precise information from the rapidity of the approximation of a signal by trigonometric polynomials (as a function of their degree), or from the decomposition of a signal into a series of polynomials \cite{5}.

Principal component analysis (PCA) seeks the best (in an $\mathcal{L}_{2}$-sense) low-rank representation of a given data matrix, and it enjoys a number of optimality
properties when the data are only mildly corrupted by small noise \cite{6}. Spectral methods hold a central place in statistical data analysis, and indeed the spectral decomposition of a positive-definite kernel underlies a variety of classical approaches, such as PCA\cite{7} usually realized using Singular Value Decomposition (SVD). Essentially, the classical SVD has associated with it the decomposition of space into the direct sum of invariant subspaces \cite{8}. But unfortunately, SVD gives linear combinations of up to all the data points, these vectors are notoriously difficult to interpret in terms of the data and processes generating the data, so CUR matrix decompositions were developed for improved data analysis \cite{9}. Stephane G. Mallat defines an orthogonal multiresolution representation
called a wavelet representation, and is computed with a pyramidal algorithm based on convolutions with quadrature mirror filters (denoted as ${\tt{qmf}}()$)\cite{4}. That is ${\tt{qmf}}(l_{1},l_{2},l_{3},l_{4},\ldots)=(l_{n},-l_{n-1},l_{n-2},-l_{n-3},\ldots)$. Independent component analysis is a framework for separating a mixture of different components into its constituents, and has been proposed for many applications \cite{10}. More recently, there exists a series of beautiful papers concerned with problem of finding the sparsest decomposition of a signal using waveforms from a highly over-complete dictionary \cite{11}.

In aforementioned achievements, many are based on matrix decomposition mainly using SVD, the difficulty of interpretation and data completion requirements make these achievements sometimes inapplicable. To tackle the difficulty, CUR matrix decomposition is produced \cite{9}; and to remove requirements, low-rank approximation of matrices with missing entries are proposed \cite{12}\cite{13}; Existing wavelet representations, such as the one proposed in \cite{4}, are usually with fixed coefficients. Thus, up to now in known researches there is no a framework straightforwardly producing general D\&R filters for signals. The framework only provides a minimal constraints for getting D\&R filters, and special requirements can be implemented by adding appropriate constraints. This paper is an explorer in this direction.

\section{the Model}
If a real data sequence $L_{d}\equiv [l_{1},l_{2},\ldots,l_{2n}]$ satisfies the following $n$ constraints
\bea \label{eq.cnt}
\left\{
  \begin{array}{ll}
    l_{1}l_{2n-1}+l_{2}l_{2n}=0\\
    l_{1}l_{2n-3}+l_{2}l_{2n-2}+l_{3}l_{2n-1}+l_{4}l_{2n}=0\\
    \ldots \\
    l_{1}l_{2n-(2n-3)}+l_{2}l_{2n-(2n-4)}+\ldots+l_{2n-2}l_{2n}=0\\
    \sum_{i=1,3,\ldots,2n-1}l_{i}=\sum_{i=2,4,\ldots,2n}l_{i}\\
        l_{1}^{2}+l_{2}^{2}+\ldots+l_{2n}^{2}=1
  \end{array}
\right.
\eea
$L_{d}$ can act as scaling function, and the associated mother function is $H_d=-{\tt{qmf}}(L_d)$. The reconstructed low-pass filter, $L_{r}$, is  $\tt{rev}(L_d)$ ( ${\tt{rev}}(l_{1},l_{2},l_{3},l_{4},\ldots)=(l_{n},l_{n-1},l_{n-2},l_{n-3},\ldots)$), and the reconstruction high-pass filter $H_{r}$ is $H_{r}={\tt{qmf}}(L_{r})$.

For instance, let $n=3$. There are $3$ constraints in \eqref{eq.cnt}, and we can  get
$l_2=\frac{-l_1l_5}{l_6}$, $l_3=\frac{(-l_1l_5/l_6+l_6-l_1-l_5)(-l_1l_5/l_6+l_6)}{(-l_1l_5/l_6+l_6+l_1+l_5)}$ and
$l_4=\frac{-(-l_1l_5/l_6+l_6-l_1-l_5)(l_1+l_5)}{(-l_1l_5/l_6+l_6+l_1+l_5)}$
with randomly specified $l_{1}, l_{5}$ and $l_{6}$. When $n$ takes 4, if $l_{1},l_{6},l_{7}$ and $l_{8}$ are specified, the remained parameters can be derived by solving a quadratic problem. Of course, some sets of $l_{1},l_{6},l_{7}$ and $l_{8}$ may make the quadratic problem have no solutions, and at this time, $L_{d}$, $H_{d}$ as well as $L_{r}$ and $H_{r}$ cannot come into being with given $l_{1},l_{6},l_{7}$ and $l_{8}$. There are formulas for solving the cubic and quartic equations, and for higher degrees, the Abel--Ruffini theorem asserts that there can not exist a general formula in radicals. Hence,
when $n$ is large (for example, $n\ge 7$), there is no explicit formula for solving $L_{d}$ coefficients from \eqref{eq.cnt} when half of $L_{d}$ parameters are given. However, root-finding algorithms (such as bracketing methods, iterative methods, etc) may be used to find numerical approximations of the roots of \eqref{eq.cnt}, and in the latter we will give an efficient iterative algorithm to solve \eqref{eq.cnt}.

Many existing wavelet filtering banks are special solutions of \eqref{eq.cnt}, and actually the equation \eqref{eq.cnt} constructs a frame consisting of many wavelet transforming sets. For instance, as is well known, the coefficients of Daubechies wavelets ``DB3" is $[0.0352,-0.0854,-0.1350,0.4599,0.8069,0.3327]$. With $l_{1}, l_{5}$ and $l_{6}$ given as before, $l_{2}\sim l_{4}$ can be solved from \eqref{eq.cnt} and are the same as given in ``DB3".  In analogy, Coiflets coefficients are all in accordance with \eqref{eq.cnt}. The fact that many existing wavelet banks have been depicted by \eqref{eq.cnt} shows that, we have many choices to implement decomposing signals. And we can impose additional constraints to rule partial coefficients of $L_{d}$. For example, when $n=3$, three coefficient $l_{1}$, $l_{5}$ and $l_{6}$ can be randomly fixed with the remains accordingly fixed. In this case, we can impose additional constraints on $l_{1}$, $l_{5}$ and $l_{6}$. For example, preconditioning $l_{1}$, $l_{5}$ and $l_{6}$ so as to make the energies distributed in components have larger differences, so the sparsity is outwardly stuck out and we can remove the trivial parts so as to make saving storage reduced.

Why the coefficients satisfying \eqref{eq.cnt} can serve as universal D\&R filtering banks? To tackle this problem, the following theorem is derived.
\begin{theorem} \label{theorem1}  When $l_{1}$, $l_{2}$, \ldots, $l_{2n}$ satisfy \eqref{eq.cnt}, the filters, $L_{d}=[l_{1},l_{2},\ldots,l_{2n}]$ and $H_{d}=-{\tt{qmf}}(L_{d})$, decompose a signal into two parts, and the two filters $L_{r}={\tt{rev}}(L_{d})$ and $H_{r}={\tt{qmf}}(L_{r})$ can reconstruct the primary signal from the two components.
\end{theorem}
 {\it {\bf Proof}}: see Appendix A. $\blacksquare$

This theorem shows that $l_{1},l_{2},\ldots,l_{2n}$ can construct the D\&R filters, provided that they form a solution of \eqref{eq.cnt}. In nature, the equation $\sum_{i=1,3,\ldots,2n-1}l_{i}=\sum_{i=2,4,\ldots,2n}l_{i}$ in \eqref{eq.cnt} comes from $H_{d}$, which extracts the high frequency part of signals through gradient operations; and $H_{d}$ can also be seen as a gradient mask whose coefficients are required to be zero summation. The other $n$ equations of \eqref{eq.cnt} comes from reconstruction requirements, and can be seen as requiring the inverse Fourier transform of the element-wise product $\mathcal{F}(L_{d})$ and $\mathcal{F}(rev(L_{d}))$ to be zero at even position except one at $2n$ position. There are $2n$ pending variables while there are only $n+1$ constraints, so \eqref{eq.cnt} has infinite solutions, which all can form $L_{d}$, $H_{d}$, $L_{r}$ and $H_{r}$. Especially, existing discrete wavelet transformations, whose filtering banks are constructed like $L_{d}$, $H_{d}$, $L_{r}$ and $H_{r}$, are all solutions of \eqref{eq.cnt}. That is to say, \eqref{eq.cnt} is a unified presentation of many discrete wavelet transformations.

\section{the Algorithm}
With the increase of $n$, getting the analytical solution of \eqref{eq.cnt} becomes difficult and impossible. So, we propose a numerical method tackle this problem, which is implemented by three steps depicted below.
\begin{description} \label{method.step}
  \item[step 1:] Randomly initialize $l_{1},l_{2},\ldots,l_{2n}$, and fix an error threshold values $\epsilon$ and iteration stop number $N$.
  \item[step 2:] For $i$ from $1$ to $2n$, let
\bea \label{eq.mtd}
l_{i}=\frac{\sum_{k=1}^{n}\alpha_{k}\beta_{k}}{\sum_{k=1}^{n}\beta_{k}^{2}},
\eea
where $\frac{\alpha_{k}}{\beta_{k}}$ for $k=1,2,\ldots,n$ will make the equations in \eqref{eq.cnt} hold except $
      l_{1}^{2}+l_{2}^{2}+\ldots+l_{2n}^{2}=1$; specially, if an equation in \eqref{eq.cnt} does not involve $l_{i}$, let the corresponding $\alpha_{k}$ and $\beta_{k}$ be zero; Normalize $L_{d}=[l_{1},l_{2},\ldots,l_{2n}]$ to make $l_{1}^{2}+l_{2}^{2}+\ldots+l_{2n}^{2}=1$ hold.
  \item[step 3:] Calculate the summation of the absolute residuals of all equations in \eqref{eq.cnt}. If the summation is less than $\epsilon$ or the iteration number achieves $N$, then exit. Otherwise, go to step 2.
\end{description}

\begin{theorem} \label{theorem2}  By {\textbf{Step 1--3}}, the numerical solution of the model \eqref{eq.cnt} can be gotten.
\end{theorem}
 {\it {\bf Proof}}: see Appendix B. $\blacksquare$

The equations in \eqref{eq.cnt} cannot definitely fix $l_{1},l_{2},\ldots,l_{2n}$. With each randomly initialized $l_{1},l_{2},\ldots,l_{2n}$, Theorem \ref{theorem2} has shown that the proposed algorithm can get a solution, and the solution can form $L_{d}$ and the accompanied filters due to Theorem \ref{theorem1}. So, Theorem \ref{theorem1} and \ref{theorem2} construct a framework to build D\&R filtering banks for discrete data sequence such as images, sampled time-varying signals. But it must be noticed that, though the solutions of \eqref{eq.cnt} include the ones satisfying the requirements imposed by the wavelet analysis, many additional constraints can be further imposed on \eqref{eq.cnt}. The proposed three steps are easy to implement, and it only provides a trick to find a number of real numbers satisfying \eqref{eq.cnt}. Some of $l_{1},l_{2},\ldots,l_{2n}$ can be prefixed, and optimizing the others can use this algorithm. But if the prefixing is not proper, more time may be required. Of course, this algorithm uses the idea of alternating minimization as used in \cite{593}.

\section{Experimental Results and Discussions}

\begin{figure}[htbp]
  \centering
  \includegraphics[width=6cm]{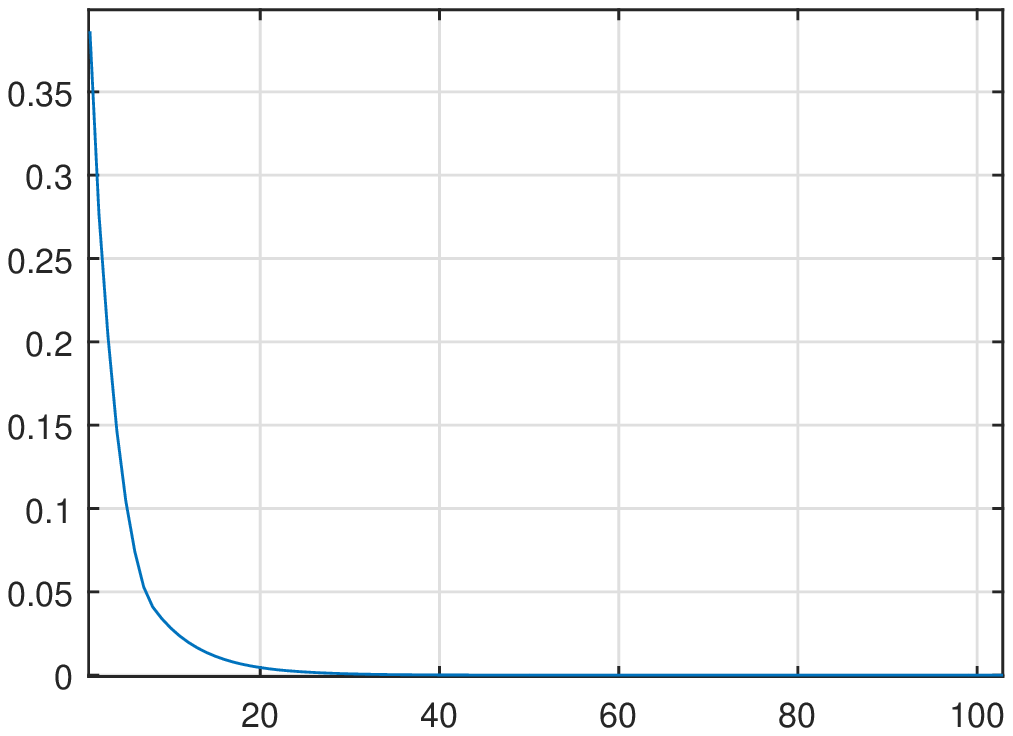}
  \caption{The convergence performance of the algorithm solving \eqref{eq.cnt} with $n=8$, and the final converged result $L_{d}$=[0.5875,-0.0583,-0.1553,0.0594,0.2736   -0.0376,-0.0432,-0.1493,-0.0068,0.4646,0.0597,0.5446,-0.0043,-0.0748,-0.0041,-0.0414].}\label{Fig.conv}
\end{figure}

To show the convergence performance of the method, let $n=8$. The Lyapunov function defined in \eqref{eq.app2.1} decreases with the iteration numbers, and the detail can refer to Figure \ref{Fig.conv}. Actually, as the Lyapunov function is a quadratic function with respect to anyone of $l_{1},l_{2},\ldots,l_{2n}$, its convergence speed is fast, and Figure \ref{Fig.conv} has demonstrated this effect. If additional constraints have imposed on \eqref{eq.cnt}, the speed may be quickened further; of course, it may become slower if the constraints are improper.

\begin{figure}[htbp]
\centering
\includegraphics[width=8cm]{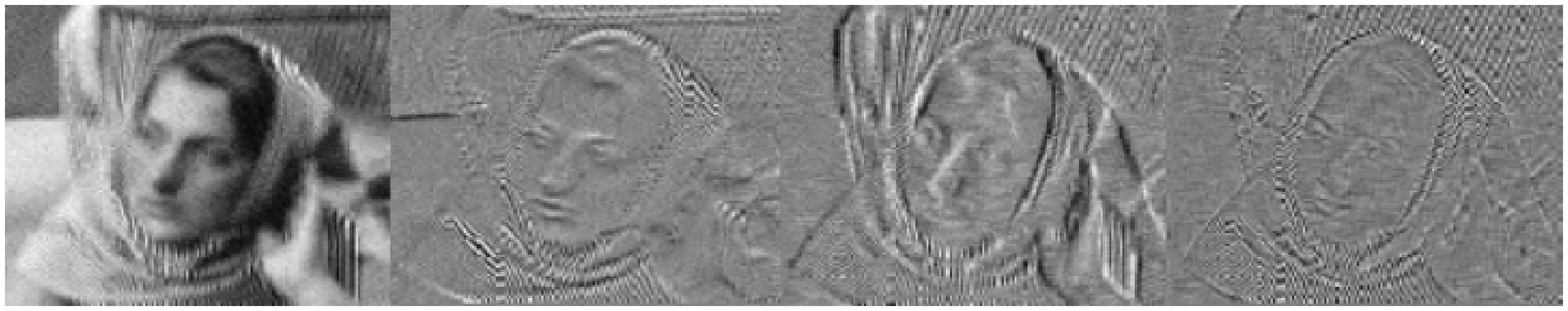}
{\scriptsize{$L_{d}=[ 0.2591,0.0343,0.5510,-0.1058,-0.1030,0.7786],\delta=$5.8265E-13}}
\includegraphics[width=8cm]{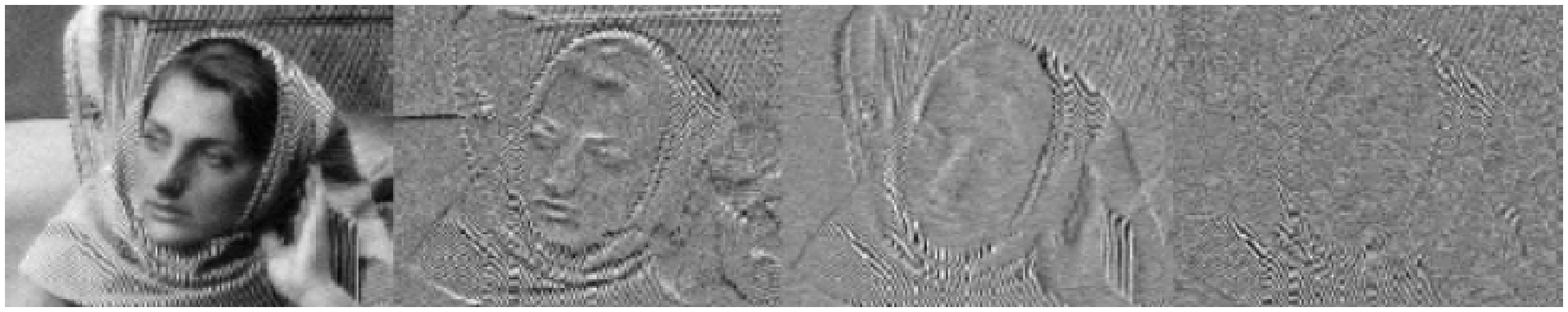}
{\scriptsize{$L_{d}=[-0.1296,0.0715,-0.1037,0.0795,-0.4739,-0.8582],\delta=$1.9895E-13}}
\includegraphics[width=8cm]{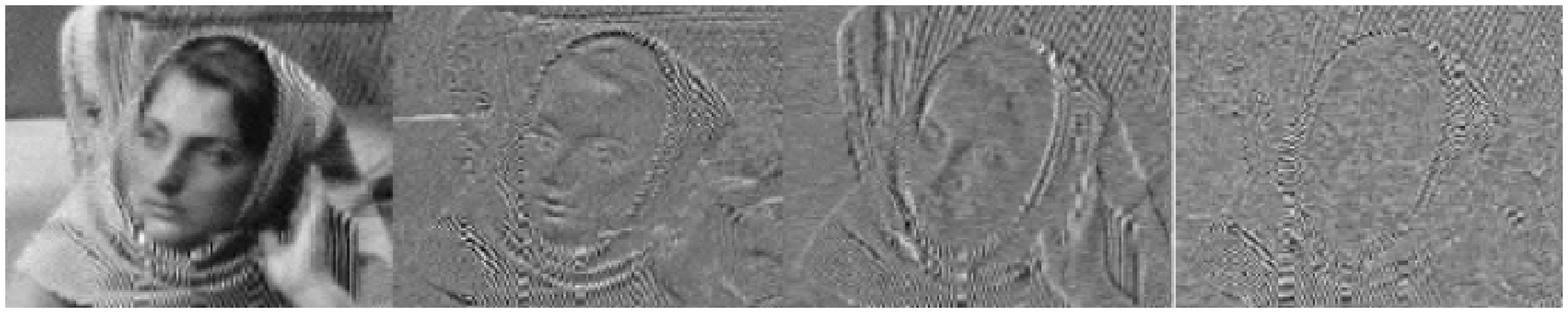}
{\scriptsize{$L_{d}=[-0.1139,0.5029,-0.0144,0.0150,0.8354,0.1892],\delta=$3.1264E-13}}
\includegraphics[width=8cm]{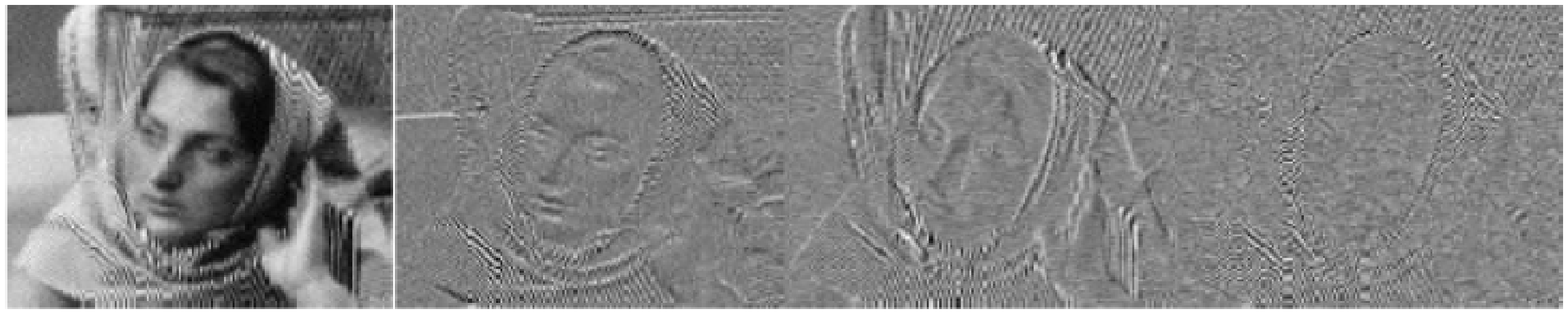}
{\scriptsize{$L_{d}=[0.3458,0.7988,-0.0759,0.0976,0.4373,-0.1893],\delta=$9.0949E-13}}
\caption{A lot of filter banks are constructed by the proposed framework, and they all can decompose signals (or images) into components. Simultaneously, the reconstruction error $\delta$ is almost zero, less than 1E-12. }\label{fig.many.L6}
\end{figure}

To show the decomposition effect and reconstruction accuracy, let $n=3$. The different filters, decomposition results and reconstruction accuracy are shown in Figure
\ref{fig.many.L6}, where each row includes 4 images corresponding to the main, horizontal, vertical and diagonal parts due to the decomposition of one image. From Figure \ref{fig.many.L6}, we can see that: 1) with the same $n$, there are a large number of filtering banks $L_{d}, H_{d},L_{r}, H_{r}$ with different coefficients; 2) and they can produce different decomposition results, the main part (the left image inset) contains most cues of the images, in each image row, the image cue seems decreasing from left and right; 3) the reconstruction accuracy $\delta$, defined as the maximal absolute difference between entries in the primary signal and the corresponding entries in the reconstructed signal, is almost zero, less than $1E-12$, that is to say, $l_{1}, l_{2},\ldots,l_{2n}$ satisfying \eqref{eq.cnt} perfectly form the D\&R filter banks.

\begin{table}[btp]
\caption{Some $L_{d}$ with different lengthes} \label{tab.1}
\end{table}
\begin{table}[h]
\centering
\begin{tabular}{c|c|c|c|c}
    n=4&n=5&n=6&n=7&n=8\\
    \hline
   0.2856&-0.1033&   -0.5898 & 0.2234 & -0.0021 \\
   0.3308&-0.3900 & -0.6356 & -0.8473 & -0.0010 \\
  -0.2345&0.1541 & 0.0314 & -0.2672 & 0.0659 \\
  0.2736&-0.1268 & -0.1777 & 0.0400 & 0.0397 \\
  0.1858&0.0538 & -0.2926 & -0.0054 & 0.0351 \\
  0.5086&-0.0284 & 0.1314 & 0.0682 & -0.2904 \\
   0.4702&-0.0284 & 0.1024 & -0.0561 & 0.0667 \\
  -0.4060&-0.3693 & 0.0530 & 0.0384 &  0.1552  \\
  -&-0.7832 & 0.1706 & -0.0983 & -0.1573 \\
  -&0.2075 & -0.1982  & 0.0357 & 0.0316 \\
  - &- & -0.1292 & -0.2041 &  -0.0071 \\
  - &- & 0.1199 & 0.0369 & 0.3433 \\
  - &- & - & -0.2995 & 0.7973 \\
  - &- & - & -0.0790 & 0.2286 \\
  - &- & - & - & -0.0913 \\
  - &- & - & - & 0.2000\\
  \hline
 \end{tabular}
\end{table}

\begin{figure}[htbp]
\centering
\includegraphics[width=8cm]{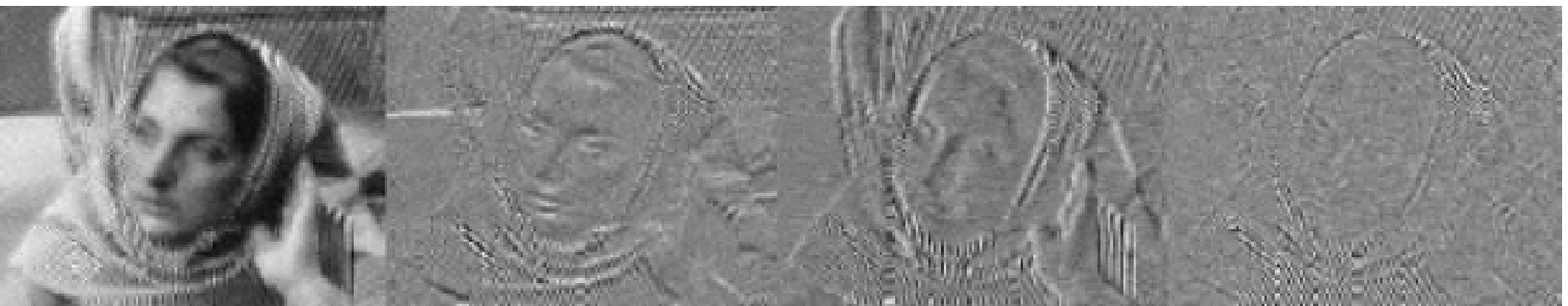}
\includegraphics[width=8cm]{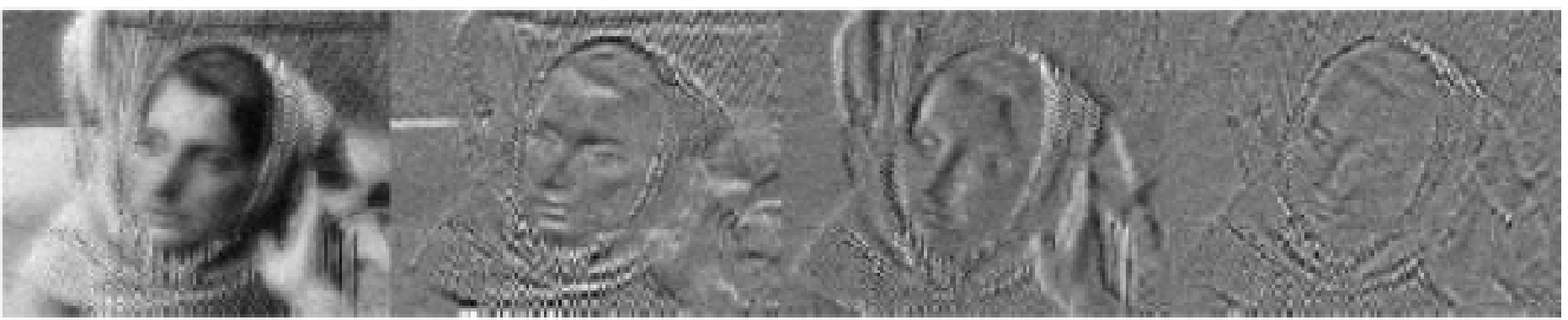}
\includegraphics[width=8cm]{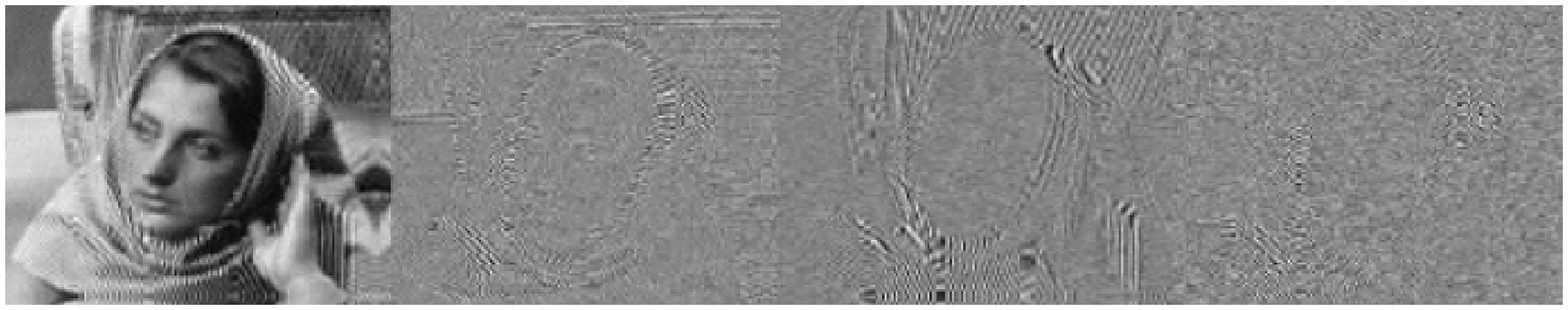}
\includegraphics[width=8cm]{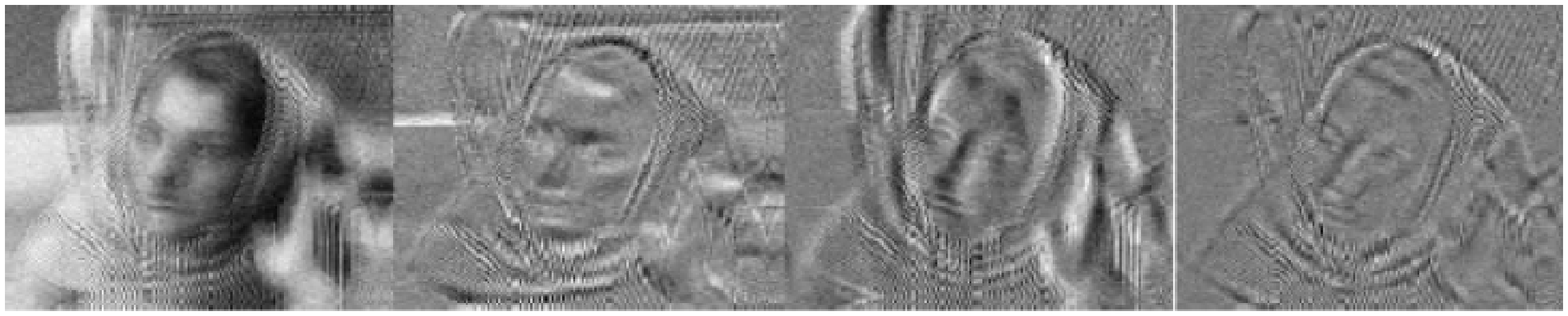}
\includegraphics[width=8cm]{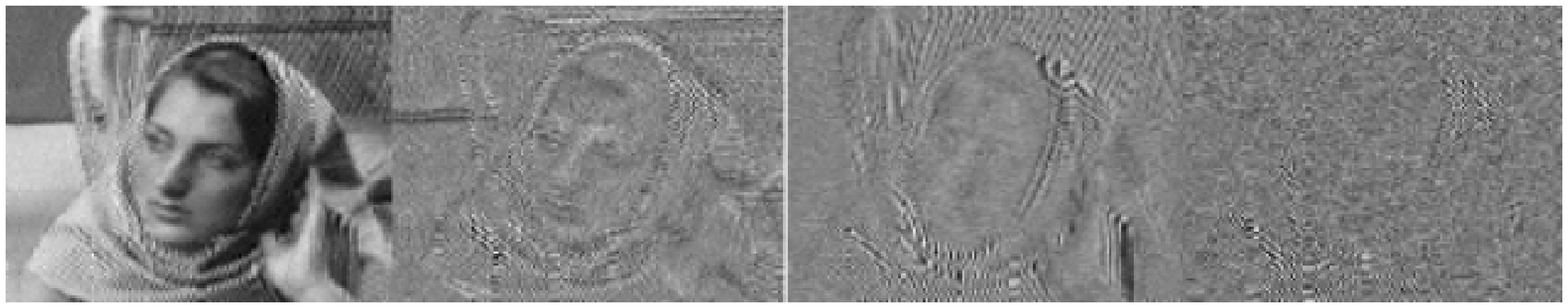}
\caption{The decomposition results of $L_{d}$ and $H_{d}$ with length $n=4,5,6,7,8$. The filter coefficients can refer to Table 1, and $\delta$=3.9790E-13, 2.8422E-13, 3.1264E-13, 4.5475E-13 and 7.1054E-13 respectively.}\label{fig.many.diff.n}
\end{figure}

To see how the change of $n$ affects the D\&R, let $n=4,5,6,7,8$. The coefficients are listed in Table \ref{tab.1}, and the reconstruction accuracy as well as the decomposition results are shown in Figure \ref{fig.many.diff.n}. From the experimental results, we can see that: 1) If more image detail is preserved in the main part, then less will be in the other parts. For example, comparing the 4th row corresponding $n=7$ with the other rows, we can see the the main part is most blurred, but the other parts contain more image information. 2) Larger $n$ does not mean better decomposition performance and reconstruction accuracy. But from \eqref{eq.cnt} we know larger $n$ means more pending parameters needed to be fixed. So, if there are more additional requirements needed to be imposed on \eqref{eq.cnt}, larger $n$ can satisfy and make the model \eqref{eq.cnt}, which has been attached with additional constraints, have solutions. 3) The decomposition results have many potential applications. For instance, in the 3rd and 5th rows, the main part contains most image information, thus the corresponding filter banks are suitable for image compression; while in the 2nd and 4th rows the vertical part contains almost whole shape features of the image object, and these filter banks may be applicable to pattern recognition area. In all, the model \eqref{eq.cnt} is a basement which can produce a mass of decomposition results, on which all kinds of tricks can be performed, and the reconstruction filtering banks are also on hand at the same time. So, by model \eqref{eq.cnt} we can realize handling signals in component or local zone, and demonstrating effect in primary domain.

\begin{figure}[htbp]
 \centering
\includegraphics[width=8cm]{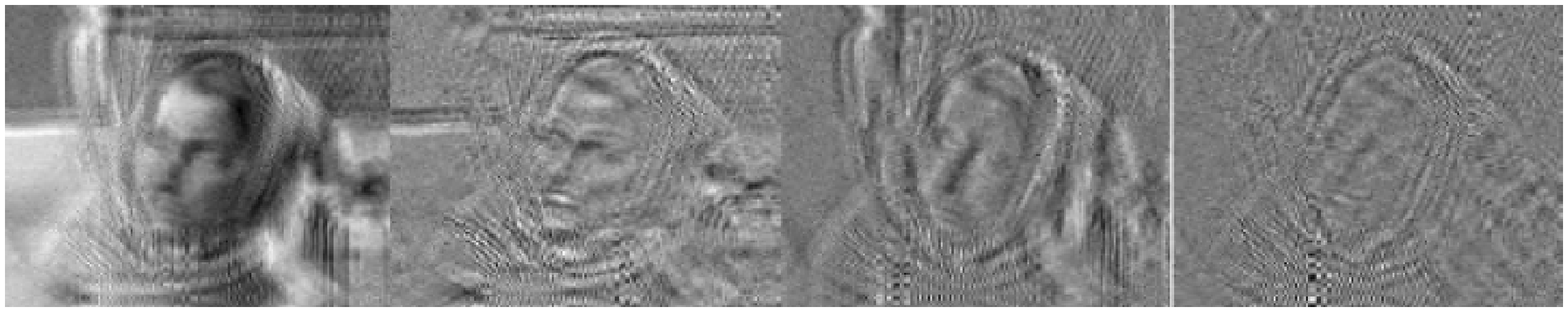}
\includegraphics[width=8cm]{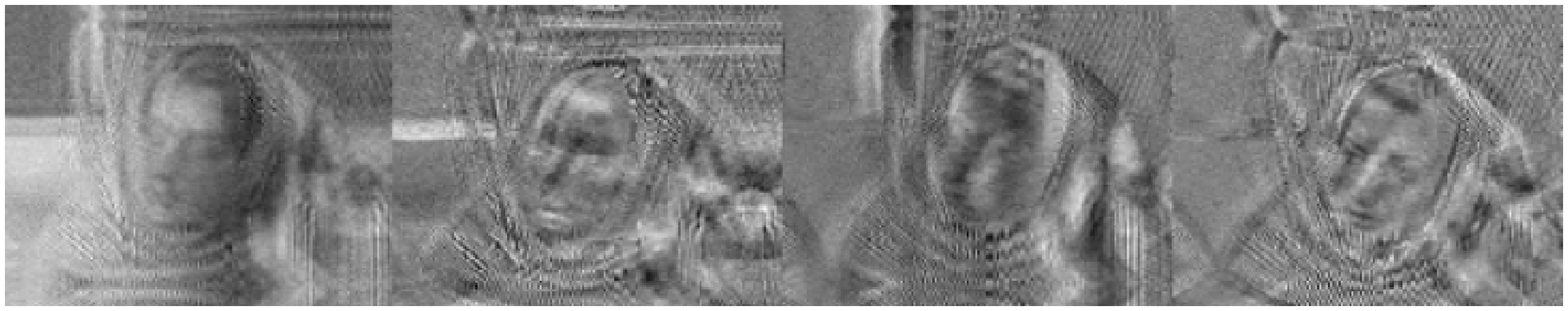}
\includegraphics[width=8cm]{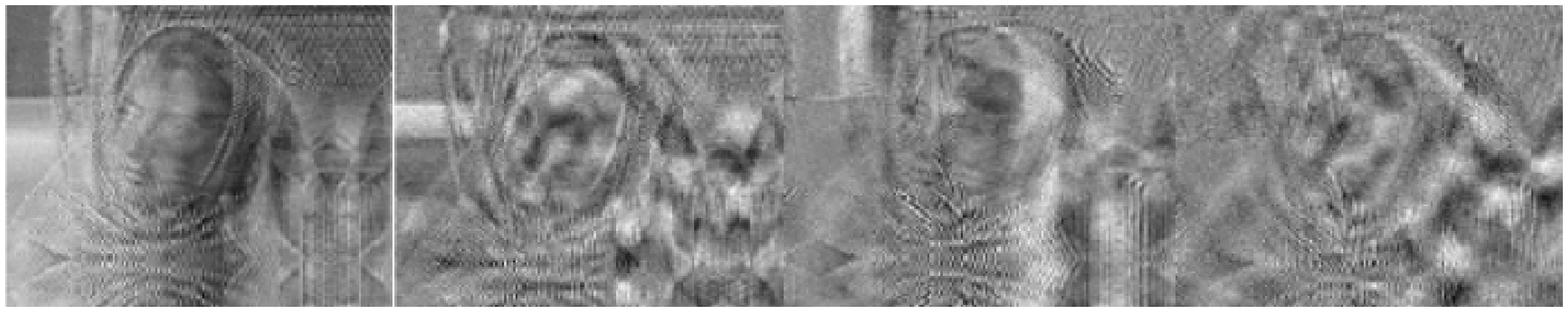}
\caption{From top to down, the filter size is 30, 40 and 50, and $\delta$=3.9790E-13,4.8317E-13,5.1159E-13 respectively.}\label{fig.many.diff.ln}
\end{figure}

As we know, $L_{d}$ and $H_{d}$ are orthogonal to each other, so are $L_{r}$ and $H_{r}$. In decomposition, continuous signals can be seen as smooth manifolds, and the discrete signals can be seen as the sampled version of the manifolds. In this case, $L_{d}$ and $H_{d}$ can be seen as two local operators performed on patches of the manifold, and the patch size is in accordance with the length of $L_{d}$ or $H_{d}$. From Riemannian geometry and Nyquist--Shannon sampling theorem, we know $L_{d}$ and $H_{d}$ with smaller size corresponds to a smaller sampling window; In other words, adhering to smaller patches can reflect more details of a manifold. From Figure \ref{fig.many.diff.ln}, we can see there is aliasing effect when filter length takes relatively large values. So, using \eqref{eq.cnt} to produce $L_{d}$ and $H_{d}$ with larger size, it is best to add constraints to improve the concentration degree of $L_{d}$ and $H_{d}$.

\section{Conclusions}
Decomposing and reconstructing discrete data sequence are widely used, and a general framework for decompositions and reconstructions are usually fundamental because special requirements can additionally imposed. But now, there is no this general framework. Thus, in this paper a model is proposed to build a general Decomposition and Reconstruction (D\&R) filtering banks. The model is composed of $n+1$ equations when the filter length is $2n$, and anyone solution of the model can act as the coefficients of the D\&R filter banks. The model is ill-posed and cannot be analytically solved when $n$ is large, so a numerical algorithm is proposed to solve the model. Noticeably, existing D\&R discrete filtering banks (such as  wavelet filtering banks called as Daubechies, Coiflets, Symlets, Meyer, etc) are solutions of the model. Many special constraints can imposed on the model, making the decomposition go along the special requirements. And some tricks can operated on the decomposed components, then using the reconstruction filters to get the special handled signals. The effectiveness of the model and the numerical algorithm is demonstrated by a large number of experimental results, and the reconstruction accuracy is excellent. In all, the proposed general D\&R framework provides an unifying fundamental for signal processing using components, and is widely applicable in practice.

\section{Appendix A: Proof of Theorem \ref{theorem1}} \label{APP1}

{\it {\bf Proof:}} Let $S=[s_{1},s_{2},\ldots,s_{m}]$ denote a real data sequence. Before $S$ is convoluted with $L_{d}=[l_{1},l_{2},\ldots,l_{2n}]$ (assuming $2n<m$ without loss of generality) and $H_{d}=-{\tt{qmf}}(L_{d})$, $S$ is circularly extended as follows
 \bea
S_{e}=[s_{2n-1},\ldots,s_{2},s_{1},s_{1},s_{2},\ldots,s_{m}].
\eea
Then the low frequency component $P=[p_{1},p_{2},\ldots]$ is gotten by sampling the convolution result, $S_{e}*L_{d}$, at even positions, and the high frequency component $Q=[q_{1},q_{2},\ldots]$ is analogously gotten from $S_{e}*H_{d}$. That is
\bea \nonumber
\begin{split}
&p_{1}=s_{2n-2}l_{2n}+s_{2n-3}l_{2n-1}+\ldots+s_{1}l_{3}+s_{1}l_{2}+s_{2}l_{1}
\\ \nonumber
&\hspace{0.3cm}=s_{2n-2}l_{2n}+s_{2n-3}l_{2n-1}+\ldots+s_{1}(l_{3}+l_{2})+s_{2}(l_{1}+l_{4}),
\\ \nonumber
&p_{2}=s_{2n-4}l_{2n}+s_{2n-5}l_{2n-1}+\ldots+s_{2}l_{3}+s_{3}l_{2}+s_{4}l_{1}
\\ \nonumber
&\hspace{0.3cm}=s_{2n-4}l_{2n}+s_{2n-5}l_{2n-1}+\ldots+s_{1}(l_{4}+l_{5})+s_{2}(l_{3}+l_{6})
\\ \nonumber
&\hspace{0.7cm}+s_{3}(l_{2}+l_{7})+s_{4}(l_{1}+l_{8}),
\\ \nonumber &\hspace{3cm} \ldots\ldots \\ \nonumber
&p_{n-1}=s_{2}l_{2n}+s_{1}l_{2n-1}+s_{1}l_{2n-2}+s_{2}l_{2n-3}+\ldots+s_{2n-2}l_{1},
\\ \nonumber
&p_{n}=s_{1}l_{2n}+s_{2}l_{2n-1}+\ldots+s_{2n}l_{1},
\\ \nonumber
&p_{n+1}=s_{3}l_{2n}+s_{4}l_{2n-1}+\ldots+s_{2n+2}l_{1},
\\ \nonumber &\hspace{3cm} \ldots\ldots \\ \nonumber
&p_{n+k}=s_{2k+1}l_{2n}+s_{2k+2}l_{2n-1}+\ldots+s_{2(k+n)}l_{1}
\end{split}
\eea
and
\bea \nonumber
\begin{split}
&q_{1}=s_{2n-2}l_{1}-s_{2n-3}l_{2}+\ldots+
\\ \nonumber
&\hspace{0.7cm}
s_{1}(l_{2n-1}-l_{2n-2})+s_{2}(l_{2n-3}-l_{2n}),
\\ \nonumber
&q_{2}=s_{2n-4}l_{1}-s_{2n-5}l_{2}+\ldots+s_{1}(l_{2n-3}-l_{2n-4})+
\\ \nonumber
&\hspace{0.7cm}s_{2}(l_{2n-5}-l_{2n-2})+s_{3}(l_{2n-1}-l_{2n-6})+s_{4}(l_{2n-7}-l_{2n}),
\\ \nonumber &\hspace{3cm} \ldots\ldots \\ \nonumber
&q_{n-1}=s_{2}l_{1}-s_{1}l_{2}+s_{1}l_{3}-s_{2}l_{4}+\ldots+s_{2n-3}l_{2n-1}-s_{2n-2}l_{2n},
\\ \nonumber
&q_{n}=s_{1}l_{1}-s_{2}l_{2}+\ldots+s_{2n-1}l_{2n-1}-s_{2n}l_{2n},
\\ \nonumber
&q_{n+1}=s_{3}l_{1}-s_{4}l_{2}+\ldots+s_{2n+1}l_{2n-1}-s_{2n+2}l_{2n},
\\ \nonumber &\hspace{3cm} \ldots\ldots \\ \nonumber
&q_{n+k}=s_{2k+1}l_{1}-s_{2k+2}l_{2}+\ldots+s_{2(k+n)-1}l_{2n-1}-s_{2(k+n)}l_{2n}.
\end{split}
\eea
The length of $P$ or $Q$ is only half of $S$. When reconstructing $S$ from $P$ and $Q$ by $L_{r}$ and $H_
{r}$, we need to make
\bea \nonumber
S=[0,p_{1},0,p_{2},0,p_{3},0\ldots]*L_{r}+[0,q_{1},0,q_{2},0,q_{3},0\ldots]*H_{r}
\eea
hold. Combing with definitions of $p_{1},p_{2},\ldots$ and $q_{1},q_{2},\ldots$, we have
\bea
\begin{split}
&s_{1}=[0,p_{1},0,p_{2},0,p_{3},0,\ldots,p_{n}][l_{1},l_{2},l_{3},\ldots,l_{2n}]^{T}+
\\ \nonumber
&\hspace{0.7cm}
[0,q_{1},0,q_{2},0,q_{3},0,\ldots,q_{n}]
\\ \nonumber
&\hspace{0.7cm}[-l_{2n},l_{2n-1},-l_{2n-2},\ldots,-l_{2},l_{1}]^{T}
\\ \nonumber
&\hspace{0.3cm}=[q_{n},p_{1},q_{n-1},p_{2},q_{n-2},p_{3},\ldots,q_{1},p_{n}][l_{1},l_{2},l_{3},\ldots,l_{2n}]^{T}
\\ \nonumber
&\hspace{0.3cm}=[l_{1},l_{2},l_{3},\ldots,l_{2n}][q_{n},p_{1},q_{n-1},p_{2},\ldots,q_{1},p_{n}]^{T},
\end{split}
\eea
that is,
\bea \label{eq.s1}
s_{1}=[l_{1},l_{2},l_{3},\ldots,l_{2n}]\mathbf{M}[s_{1},s_{2},\ldots,s_{2n}]^{T}
\eea
where
\bea \nonumber
\mathbf{M}\triangleq\left[
\begin{array}{l}
    l_{1},-l_{2},l_{3},-l_{4},\ldots, l_{2n-1},-l_{2n} \\
    l_{3}+l_{2},l_{1}+l_{4},\ldots,l_{2n-1},l_{2n},0,0\\
    l_{3}-l_{2},l_{1}-l_{4},l_{5},\ldots,l_{2n-1},-l_{2n},0,0\\
    l_{4}+l_{5},l_{3}+l_{6},l_{2}+l_{7},l_{1}+l_{8},\ldots,l_{2n-1},l_{2n},0,0,0,0\\
   \vdots\\
    l_{2n-1}-l_{2n-2},l_{2n-3}-l_{2n},-l_{2n-4},\ldots,-l_{2},l_{1},0,0\\
    l_{2n},l_{2n-1},\ldots,l_{2},l_{1}
  \end{array}
\right].
\eea
The above formula will hold if the following relation hold,
\bea
\begin{split}
&[1,0,\ldots,0]=[l_{1},l_{2},l_{3},\ldots,l_{2n}]\mathbf{M}
\end{split}
\eea
and this comes down to the following $2n$ conditions with each one produced by multiplying $[l_{1},l_{2},l_{3},\ldots,l_{2n}]$ with the $i$th column of $\mathbf{M}$
\bea \label{eq.app.x1.1}
\begin{split}
&1=[l_{1},l_{2},l_{3},\ldots,l_{2n}]
\\ \nonumber
&\hspace{0.3cm}[l_{1},l_{3}+l_{2},l_{3}-l_{2},l_{4}+l_{5},\ldots,l_{2n-1}-l_{2n-2},l_{2n}]^{T}
\\ \nonumber&\hspace{0.2cm}=l_{1}^{2}+l_{2}^{2}+\ldots+l_{2n}^{2},
\\ \nonumber&0=[l_{1},l_{2},l_{3},\ldots,l_{2n}]
\\ \nonumber
&\hspace{0.3cm}[-l_{2},l_{1}+l_{4},l_{1}-l_{4},l_{3}+l_{6},\ldots,l_{2n-3}-l_{2n},l_{2n-1}]^{T}
\\ \nonumber&\hspace{0.2cm}=-l_{1}l_{2}+l_{2}(l_{1}+l_{4})+l_{3}(l_{1}-l_{4})+l_{4}(l_{3}+l_{6})+
l_{5}(l_{3}-l_{6})
\\ \nonumber
&\hspace{0.5cm}+l_{6}(l_{5}+l_{8})+\ldots+l_{2n-3}(l_{2n-5}-l_{2n-2})+
\\ \nonumber
&\hspace{0.5cm}l_{2n-2}(l_{2n-3}+l_{2n})+
l_{2n-1}(l_{2n-3}-l_{2n})+l_{2n}l_{2n-1}\\
\nonumber&\hspace{0.2cm}=l_{2}l_{4}+l_{1}l_{3}+l_{4}l_{6}+l_{5}l_{3}+l_{6}l_{8}+\ldots+
\\ \nonumber
&\hspace{0.5cm}l_{2n-3}l_{2n-5}+l_{2n-2}l_{2n}+l_{2n-1}l_{2n-3}
\\\nonumber&\hspace{0.2cm}=l_{1}l_{3}+l_{2}l_{4}+l_{3}l_{5}+l_{4}l_{6}+\ldots+l_{2n-3}l_{2n-1}+l_{2n-2}l_{2n},
\\ \nonumber   &\hspace{3cm}\ldots
\\ \nonumber&
0=l_{1}l_{2n-1}+l_{2}\times 0+\ldots+l_{2n-1}\times 0+l_{2n}l_{2}
\\ \nonumber&\hspace{0.2cm}=l_{1}l_{2n-1}+l_{2n}l_{2},
\\ \nonumber&0=-l_{1}l_{2n}+l_{2}\times 0+\ldots+l_{2n-1}\times 0+l_{1}l_{2n}.
\end{split}
\eea
In above formulas, the $1$st condition is to normalize $L_{d}$; the $2$nd and $3$rd conditions are the same, that is, $l_{1}l_{3}+l_{2}l_{4}+l_{3}l_{5}+l_{4}l_{6}+\ldots+l_{2n-3}l_{2n-1}+l_{2n-2}l_{2n}=0$; for $i=2,\ldots,n-1$, the $(2i)$th and $(2i+1)$th conditions are same, $l_{1}l_{2i+1}+l_{2}l_{2i+2}+\ldots+l_{2n-2i}l_{2n}=0$; and the final condition is an identity shown as above. So, in total there are $n$ constraints on $l_{1},l_{2},\ldots,l_{2n}$. But the first condition is to normalize $L_{d}$, and this condition is required naturally with the goal of not introducing additional energy into the signal $S$. On the other side, the goal of $H_{d}*S$ is to get the high frequency component of $S$, actually performing a weighted gradient operation. So we need to make the summation of $H_{d}=-{\tt{qmf}}(L_{d})$ equals to $0$, and like discrete wavelet filter banks we impose the condition, $l_{1}+l_{3}+\ldots+l_{2n-1}=l_{2}+l_{4}+\ldots+l_{2n}$. In all, \eqref{eq.cnt} will make \eqref{eq.s1} hold.

Like the operations on $s_{1}$, we can get
\bea
\begin{split}
&s_{2}=[p_{1},0,p_{2},0,p_{3},0,\ldots,p_{n},0][l_{1},l_{2},l_{3},\ldots,l_{2n}]^{T}+
\\ \nonumber
&\hspace{0cm}
[q_{1},0,q_{2},0,q_{3},0,\ldots,q_{n},0][-l_{2n},l_{2n-1},-l_{2n-2},\ldots,-l_{2},l_{1}]^{T}
\\ \nonumber
&\hspace{0.3cm}=[l_{1},l_{2},l_{3},\ldots,l_{2n}][p_{1},-q_{n},p_{2},-q_{n-1},\ldots,q_{n},-q_{1}]^{T},
\end{split}
\eea
\bea
\begin{split}
&s_{3}=[0,p_{2},0,p_{3},0,\ldots,p_{n},0,p_{n+1}][l_{1},l_{2},l_{3},\ldots,l_{2n}]^{T}+
\\ \nonumber
&\hspace{0cm}
[0,q_{2},0,q_{3},0,\ldots,q_{n},0,q_{n+1}][-l_{2n},l_{2n-1},-l_{2n-2},\ldots,-l_{2},l_{1}]^{T}
\\ \nonumber
&\hspace{0.3cm}=[l_{1},l_{2},l_{3},\ldots,l_{2n}][q_{n+1},p_{2},q_{n},p_{3},\ldots,q_{2},p_{n+1}]^{T},
\end{split}
\eea
\bea
\begin{split}
&s_{4}=[l_{1},l_{2},l_{3},\ldots,l_{2n}][p_{2},-q_{n+1},p_{3},-q_{n}\ldots,p_{n+1},-q_{2}]^{T},
\\ \nonumber
&\hspace{3cm}\ldots
\end{split}
\eea
\bea
\begin{split}
&s_{2k-1}=[l_{1},l_{2},l_{3},\ldots,l_{2n}]
\\ \nonumber
&\hspace{0.7cm}[q_{n+k-1},p_{k},q_{n+k-2},p_{k+1},\ldots,q_{k},p_{n+k-1}]^{T},
\\ \nonumber
&s_{2k}=[l_{1},l_{2},l_{3},\ldots,l_{2n}]
\\ \nonumber
&\hspace{0.7cm}
[p_{k},-q_{n+k-1},p_{k+1},-q_{n+k-2}\ldots,p_{n+k-1},-q_{k}]^{T},
\end{split}
\eea
Substituting $p_{i}$ and $q_{i}$ into above formulas, and like the operations on \eqref{eq.s1} we can derive the constraint equation set \eqref{eq.cnt} from each one of above formulas.  So, Theorem \ref{theorem1}
holds. $\blacksquare$

\section{Appendix B: Proof of Theorem \ref{theorem2}} \label{APP2}
{\it {\bf Proof:}} First $l_{1},l_{2},\ldots,l_{2n}$ are randomly initialized. Then, how to tune $l_{k}$ for $k=1,2,\ldots,2n$ in turn when seeing the other $2n-1$ coefficient $l_{i}$ for $i\neq k$ as known. In \eqref{eq.cnt}, the first $n$ equations have no relations with the final one. So, when solving \eqref{eq.cnt}, first we consider the first $n$ equations, then the final one. With the fist $n$ equations, we build a Lyapubov functional as follows
\bea \label{eq.app2.1}
\begin{split}
&\mathcal{L}(l_{1},l_{2},\ldots,l_{2n})=(l_{1}l_{2n-1}+l_{2}l_{2n})^2+
\\ \nonumber
&\hspace{1cm}(l_{1}l_{2n-3}+l_{2}l_{2n-2}+l_{3}l_{2n-1}+l_{4}l_{2n})^2+\ldots
\\
    &\hspace{1cm}(l_{1}l_{2n-(2n-3)}+l_{2}l_{2n-(2n-4)}+\ldots+l_{2n-2}l_{2n})^2
\\ \nonumber
    &\hspace{1cm}\left(\sum_{i=1,3,\ldots,2n-1}l_{i}-\sum_{i=2,4,\ldots,2n}l_{i}\right)^2.
\end{split}
\eea
As we know, if tuning each of $l_{1},l_{2},\ldots,l_{2n}$ makes $\mathcal{L}(l_{1},l_{2},\ldots,l_{2n})$ minimized with respect to the tuned coefficient, then $\mathcal{L}(l_{1},l_{2},\ldots,l_{2n})$ will be finally minimized. From definition of $\mathcal{L}(l_{1},l_{2},\ldots,l_{2n})$ we know, when seeing $l_{k}$ as a variable and the other coefficients as constants, $\mathcal{L}(l_{1},l_{2},\ldots,l_{2n})$ is a strict convex quadratic function with respect to $l_{k}$. So, $\mathcal{L}(l_{1},l_{2},\ldots,l_{2n})$ is minimized when
\bea
\frac{\partial \mathcal{L}(l_{1},l_{2},\ldots,l_{2n})}{\partial l_{k}}=0,
\eea
which results in
\bea \nonumber
[\alpha_{1},\alpha_{2},\ldots,\alpha_{n}]^{T}[\alpha_{1},\alpha_{2},\ldots,\alpha_{n}]l_{k}=
\\ \nonumber
[\alpha_{1},\alpha_{2},\ldots,\alpha_{n}]^{T}[\beta_{1},\beta_{2},\ldots,\beta_{n}],
\eea
and this equation equals to \eqref{eq.mtd}. So, each tuning of $l_{1},l_{2},\ldots,l_{2n}$ will decrease $\mathcal{L}(l_{1},l_{2},\ldots,l_{2n})$, and the given algorithm can get numerical solutions of  \eqref{eq.cnt}. $\blacksquare$
\begin{acknowledgments}
This work is supported by NSFC under Grant 61571313 and U1633126, by funding from Sichuan Province under Grant 18GJHZ0138, and by funding under 2016CDLZ-G02-SCU from Sichuan University and Lu-Zhou city
\end{acknowledgments}


\end{article}

\end{document}